\documentclass[a4paper,12pt]{iopart}
\usepackage[cp1250]{inputenc}
\usepackage[]{amssymb}
\usepackage[]{amsthm}
\usepackage[]{mathrsfs}
\usepackage[]{tensor}
\usepackage[]{eucal}
\usepackage[symbol]{footmisc}
\usepackage{graphicx}
\usepackage{color}
\usepackage{hyperref}

\begin{document}

\title[\scriptsize{Hopping Electron Transport in Doped Polyaniline: an Experimental Verification of the FTS Model}]{Hopping Electron Transport in Doped Polyaniline: an Experimental Verification of the Fogler-Teber-Shklovskii Model}

\vspace{50pt}

\author{Mirko Ba\'{c}ani$^1$, Mario Novak$^1$\footnote{Present address: Department of Quantum Functional Materials, ISIR, Osaka University, Japan}, Ivan Kokanovi\'{c}$^1$, Dinko Babi\'{c}$^2$}
\vspace{40pt}

\address{$^1$Department of Physics, Faculty of Science, University of Zagreb, Bijeni\v{c}ka cesta 32, HR-10000 Zagreb, Croatia}
\bigskip
\address{$^2$Institute for Medical Research and Occupational Health, Ksaverska cesta 2, HR-10000 Zagreb, Croatia}
\vspace{30pt}
\ead{mbacani@phy.hr}

%\date{\today}

\vspace{40pt}

\begin{abstract}
%
%\noindent
We present a study of the temperature ($T$) dependence of the dc electrical conductivity of polyaniline pellets doped over a wide range. A crossover between low-$T$ variable-range hopping and high-$T$ nearest-neighbor hopping has been found below room temperature for most of the samples, which is atypical for conducting polymers. This allows us to construct an experimental diagram that accounts for different regimes in the hopping electron transport and closely resembles predictions of the Fogler-Teber-Shklovskii model in three dimensions.
\end{abstract}

%72.20.Ee   Mobility edges; hopping transport
%72.80.Le   Conductivity of polymers
%77.84.Jd   Dielectric materials: Polymers; organic compounds
%81.05.Lg		Specific materials: fabrication, treatment, testing, and analysis : Polymers and plastics
%81.40.Rs 	Electrical and magnetic properties related to treatment conditions
%81.40.-z 	Treatment of materials and its effects on microstructure, nanostructure, and properties
%%82.35.Cd   Polymers: properties; reactions; polymerization - conducting polymers
%72.15.Rn 	Localization effects (Anderson or weak localization) 

%\keywords{}
\maketitle

Particle transport through a disordered potential consisting of deep, localizing potential wells has for decades been puzzling. This transport usually occurs via thermally assisted hopping from one localized site to another, good examples being electron transport in strongly disordered conductors \cite{Mott1979}, and vortex creep in type-II superconductors \cite{Blatter1994}. If the well depth varies in space strongly, a particle hops to a site determined by the most favorable balance between the hop length and the energy difference between the localized states involved. When thermal activation is strong enough to diminish the importance of the spatially varying well depth, the above mechanism - called variable-range hopping (VRH) - is replaced by nearest-neighbor hopping (NNH). The most suitable observable for investigating these phenomena is electrical conductivity, which is a linear-response quantity that can contain enough information on the nature of a hopping electron transport (HET). This approach has been dominant ever since the early works of Anderson \cite{Anderson1958} and Mott \cite{Mott1969} on localized electron states in solids and their thermally assisted delocalization. Mott argued \cite{Mott1979} that HET electrical conductivity $\sigma$ generally depends on $T$ as
\begin{equation}\label{equation1}
\sigma\left(T\right) =  \eta_{\alpha} \mathrm{exp}\left[-\left(T_\alpha/T\right)^\alpha\right],
\end{equation}
\noindent
where $\eta_{\alpha}$, $T_\alpha$, and $\alpha\leq1$ contain information on the underlying microscopic mechanisms. These parameters may depend on dimensionality ($d$), interactions relevant for the transport, disorder level, localization length of electrons, etc. For VRH, $\alpha<1$, whereas NNH obeys the Arrhenius law ($\alpha=1$). 

\bigskip

VRH of noninteracting electrons results in the Mott law $\alpha=\left(1+d\right)^{-1}$. Coulomb interaction leads to an $\left|E-E_F\right|^\mu$ energy ($E$) dependence of the electron density of states (DOS) close to the Fermi energy $E_F$, with $\mu$ being a constant. This opens a soft (Coulomb) gap in the DOS, and results in the Efros-Shklovskii (ES) law where $\alpha=1/2$ is universal \cite{Efros1975,Shklovskii1984}. The above models address isotropic systems, e.g., disordered semiconductors. However, HET occurs in other materials as well, of which we pay attention to conducting polymers. For many of them, Eq. (\ref{equation1}) accounts for experimental data remarkably well \cite{KaiserRPP}, with $\alpha=1/4$ and $\alpha=1/2$ appearing often. Other $\alpha<1$ have been found as well, including $\alpha=2/5$ \cite{Aleshin1997,Chaudhuri2006,Maji2007,mario,cca} that cannot be explained neither by the traditional models (Mott, ES) nor by similar approaches \cite{Sheng1973, Wang1991, Wang1992, Zuppiroli1994}. NNH is, on the other hand, much less common in conducting polymers, and $\alpha=1$ is observed \cite{jncs,Bufon2007,Ghosh2001} infrequently, probably because conditions for its occurrence are seldom established at $T$ lower than the material degradation temperature.

\bigskip

A model with a potential to account for the complexity of $\sigma(T)$ in conducting polymers was developed by Fogler, Teber and Shklovskii (FTS) \cite{fts}. They studied the HET in coupled chain-like conductors in the presence of disorder-dependent Coulomb interaction potential that affected DOS similarly as in the ES model. The FTS model predicts $\alpha=\left(1+\mu\right)/\left(1+\mu+d\right)$, which for strongly coupled chains ($d=3$) and $\mu=0,1,2$ leads to $\alpha= 1/4, 2/5, 1/2$, respectively. At low $T$, highly disordered samples should exhibit $\alpha=1/2$, and more ordered samples $\alpha=2/5$. In the former case, the transition to NNH as $T$ increases should be direct, i.e., $\alpha=1/2\to1$, whereas in the latter case, one should observe $\alpha=2/5\to1/4\to1$. On the basis of these considerations, FTS proposed a $d=3$ diagram where crossover temperatures between different $\alpha$ values were plotted vs suitably parameterized disorder level. A support to the FTS predictions has come from recent experiments on VRH in polyaniline (PANI) \cite{mario,jncs} which is a well-known polymer that becomes conducting when doped via protonation by an acid \cite{chiang18,hRMP}. In HCl-doped samples (PANI-HCl), $\alpha$  exhibited crossovers between 1/4, 2/5, and 1/2 \cite{mario}, but the transition from VRH to NNH below room temperature (RT) was identified only occasionally \cite{jncs}. Hence, the FTS diagram - which is in many respects central to the model - could not have been constructed from the experimental data. A similar VRH behavior has been found in samples of PANI-DBSA, i.e., PANI doped with dodecylbenzenesulfonic acid (DBSA) \cite{cca}. In this case, however, VRH-to-NNH transition occurs well below RT over a wide doping range, and this opens a possibility to explore the viability of the FTS diagram.

\bigskip

In this paper, we present a systematic study of the VRH and NNH in pressed PANI-DBSA pellets doped over a wide range. For the first time, we construct the FTS diagram from experimental data and confirm its correctness. Data for those of our PANI-HCl samples which show NNH below RT supplement the diagram nicely. Hence, our findings may have consequences for general viewpoint on the HET in conducting polymers and similar conductors.

\bigskip

Detailed information on the sample synthesis, doping, and characterization (nuclear spectroscopy for elemental analysis, X-ray diffraction for structure, magnetic susceptibility) can be found in \cite{eqdop}, and here we list the main conclusions. $Q=4n_{\mathrm{S}}/n_{\mathrm{N}}$, where $n_{\mathrm{S}}$ and  $n_{\mathrm{N}}$ are concentrations of sulfur and nitrogen in PANI-DBSA, accounts for the doping level and  can be as high as 3.39 (while $Q=2$ corresponds to the complete protonation). This means that the amount of DBSA in a sample can exceed that of the protonation sites on PANI molecules. Measured transport and magnetic properties imply that all DBSA participates in the protonation when $Q\leq2$, whereas $Q>2$ means that PANI is completely protonated and extra DBSA is captured between the protonated PANI molecules. For all $Q$, the error of which is at the level of a few percent, no signature of crystalline ordering is found. $\sigma$ of the samples increases monotonically with increasing $Q$ over the whole doping range. While $\sigma\left(Q\right)$ for $Q\leq2$ is a straightforward consequence of the protonation, the increase at $Q>2$ has been attributed to a rearrangement of the fully protonated PANI molecules into a configuration where interchain electron hops are enhanced \cite{cca}. Dc conductivity measurements were carried out in a closed-cycle refrigerator from the lowest achievable temperature of $\sim10\  \mathrm{K}$ up to RT.

\begin{figure} [t]
\begin{center}
\includegraphics[width=95mm]{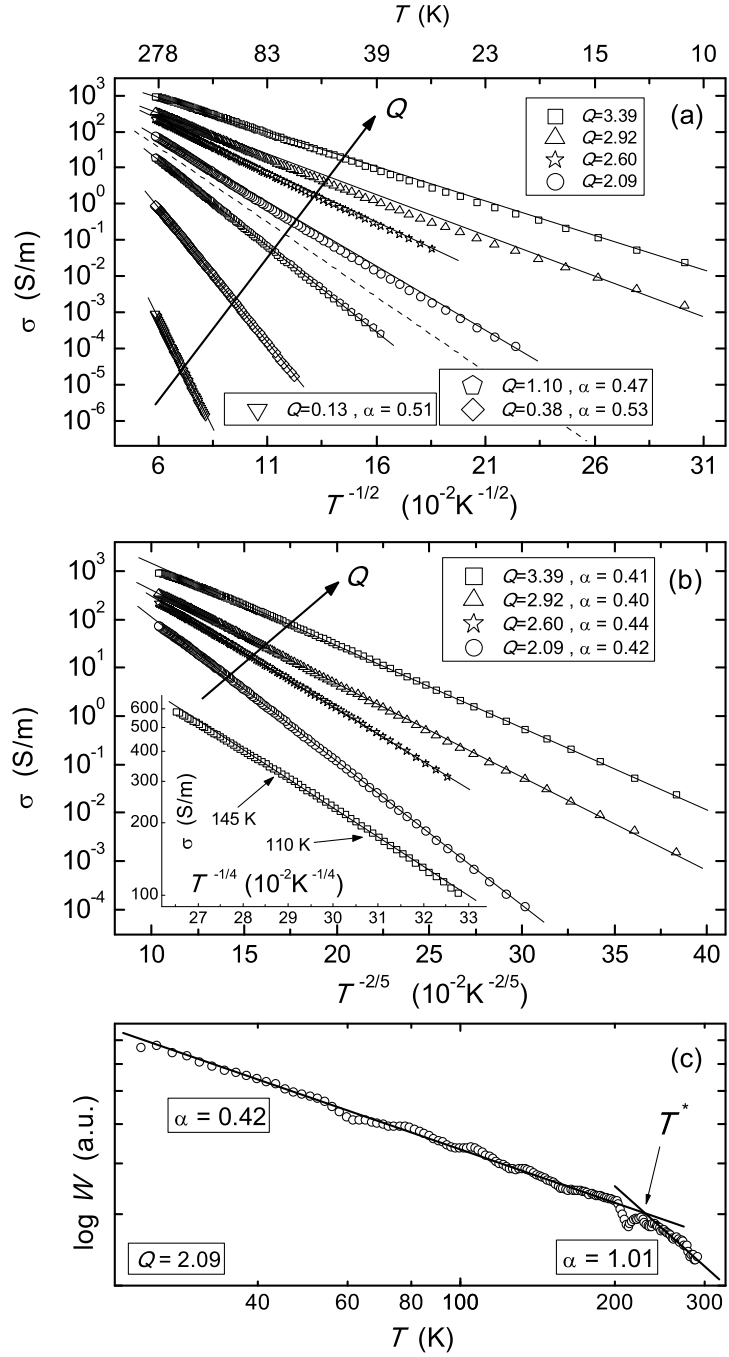}
\caption{\label{fig1}(a) Experimental $\log \sigma$ vs $T^{-1/2}$ (symbols). The solid lines are $\left[\textrm{here and in (b)}\right]$ guides to the eye.  (b) Data above the dashed line in (a), plotted as $\log \sigma$ vs $T^{-2/5}$ (symbols). Inset to (b): $\log \sigma$ vs $T^{-1/4}$ for $Q=3.39$, exhibiting linearity for $110 \lesssim T \lesssim 145$ K. The corresponding $\alpha$ determined numerically $\left[\textrm{as exemplified in (c)}\right]$ is $0.27$. (c) $\log W$ vs $\log T$ of a selected sample (symbols), displaying two distinct ranges of linearity with a crossover at $T^*$. The solid lines represent linear fits that yield $\alpha$ numerically. $\alpha$ resulting from this procedure is listed in the legends to (a) and (b).}
\end{center}
\end{figure}

\bigskip

Experimental $\sigma(T)$ curves for the obtained range of $Q$ are in Fig.~\ref{fig1}(a) plotted by the symbols as $\log \sigma$ vs $T^{-1/2}$, which provides good insight into low-$T$ data. Below the dashed line, the plots are obviously straight. Above the dashed line, the symbols and the corresponding straight solid lines (guides to the eye) do not match well, hence there is no linear dependence. These data are in Fig.~\ref{fig1}(b) plotted by the symbols as $\log \sigma$ vs $T^{-2/5}$, which results in linear plots. From Figs.~\ref{fig1}(a) and \ref{fig1}(b), we can conclude that Eq. (\ref{equation1}) accounts for low-$T$ data adequately, with $\alpha = 1/2$ for low $Q$, and $\alpha = 2/5$ for high $Q$. A similar behavior was found for PANI-HCl (in which $Q=2$ cannot be exceeded \footnote{For PANI-HCl, $Q$ is calculated by replacing $n_\textrm{S}$ with	Cl concentration $n_\textrm{Cl}$. In fully dry samples, no extra HCl can be intercalated between PANI chains.}) but the change from $\alpha=1/2$ to $\alpha=2/5$ occurred at a $Q$ lower than here \cite{mario}.

\bigskip

It is difficult to note in Figs.~\ref{fig1}(a) and \ref{fig1}(b) that the above low-$T$ values of $\alpha$ do not hold for the high-$T$ data in narrow ranges on the left-hand side of the plots. As we show later, the proper exponent in this case is $\alpha = 1$ irrespective of the low-$T$ value, indicating that NNH is the mechanism that underlies the HET in this regime. Moreover, for the sample with the highest $Q=3.39$, there is a narrow $T$ range ($\sim 110 - 145$ K) where $\alpha=1/4$, as shown in the inset to Fig.~\ref{fig1}(b). Therefore, as $T$ increases, $\alpha$ of this sample changes as $2/5\to1/4\to1$, in contrast to the other samples where this change is either $1/2\to1$ or $2/5\to1$.

\bigskip

Although $\log \sigma$ vs $T^{-\alpha}$ plots are indicative with regard to $\alpha$, we carried out an alternative analysis as well. Application of $W=\partial \ln \sigma/ \partial \ln T$ to Eq.~(\ref{equation1}) leads to $-\alpha$ being the slope in the resulting linear $\log W$ vs $\log T$ dependence, which enables a more precise extraction of $\alpha$ from experimental $\sigma (T)$. This is demonstrated in Fig.~\ref{fig1}(c) for a selected sample. Two distinct  ranges of linearity are discernible, corresponding to VRH ($\alpha=0.42$) and NNH ($\alpha=1.01$), as indicated (for all our data, error in thusly determined $\alpha$ is $\lesssim 10\%$). $T^*$ is a crossover temperature between the VRH and NNH. Crossover at $T^*$ is generally not sharp, and the $T$ range between regimes of well defined low-$T$ and high-$T$ values of $\alpha$ can be up to $\sim40$~K wide. A large crossover width can be an obstacle for identifying $\alpha=1/4$, because this VRH behavior should - according to the FTS model - take place only in a narrow $T$ range below $T^*$. Our result for the $Q=3.39$ sample, where $\alpha=1/4$ appears only over 35 K below $T^*$, supports this prediction. Moreover, we cannot rule out that $\alpha=1/4$ in some other samples might be hidden due to a large width of the transition at $T^*$.  

\begin{figure}
\begin{center}
\includegraphics[width=95mm]{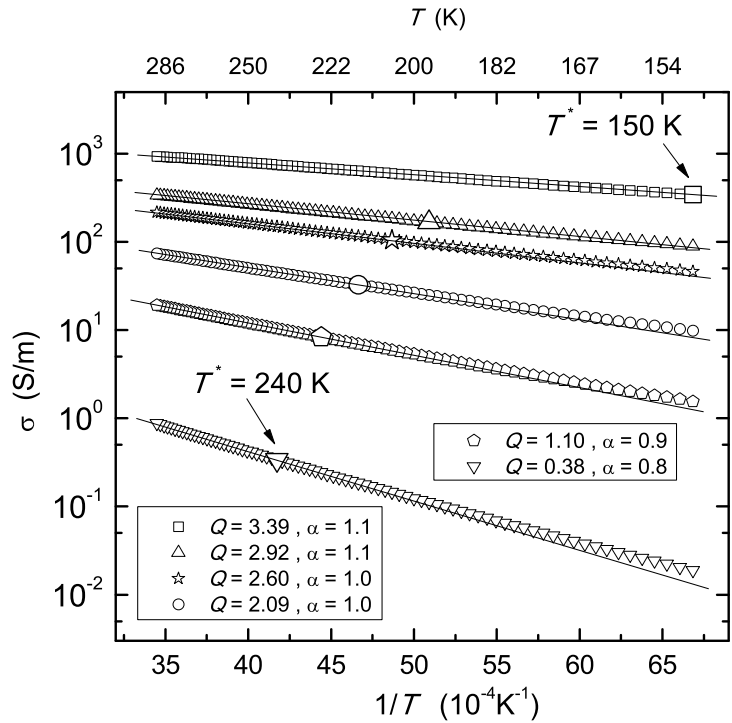}
\caption{\label{fig2}Experimental $\log \sigma$ vs $T^{-1}$ (symbols) at $Q\geq0.38$, for which $T^*$ is below RT. The solid lines are guides to the eye. $T^*$ for each sample is marked by a larger symbol. Numerical values of $\alpha$ extracted numerically are listed in the legends.}
\end{center}
\end{figure}

\bigskip

In Fig.~\ref{fig2}, we use $\log \sigma$ vs $T^{-1}$ plots to show the high-$T$ part of $\sigma(T)$ at $Q\geq0.38$, for which $T^*$ is lower than RT. Larger symbols mark $T^*$ which decreases with increasing $Q$. We also note that magnetic susceptibility $\chi$  of the samples undergoes a significant change \footnote{M. Ba\'{c}ani, I. Kokanovi\'{c}, M. Novak, and D. Babi\'{c},	unpublished}, similar to that described in \cite{epjb} for PANI-HCl, at basically the same $T^*(Q)$ as here. The fact that $T^*$ leaves a signature both in $\sigma$  and  $\chi$ implies a qualitative change in the electronic system at $T^*$.

\bigskip

\begin{figure} [h]
\begin{center}
\includegraphics[width=102mm]{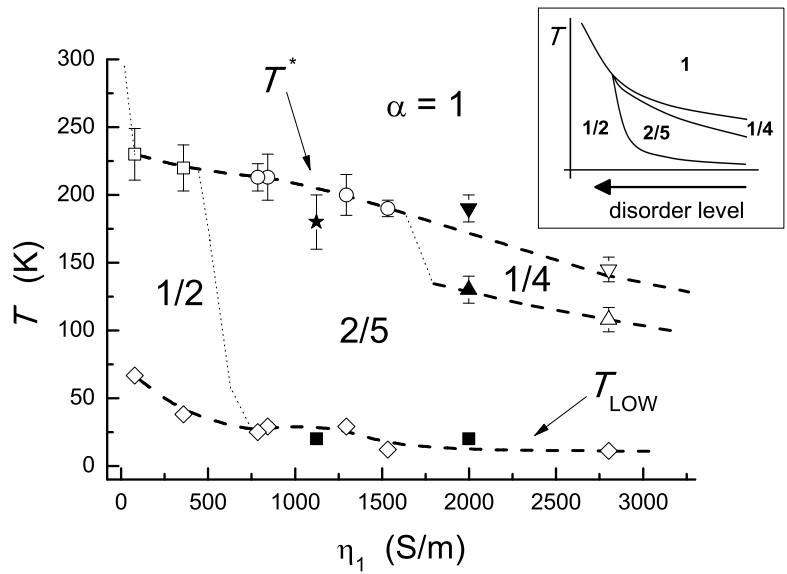}
\caption{\label{fig3}Areas of different values of $\alpha$, shown in a $T$ vs $\eta_1$ plane. The open symbols correspond to the PANI-DBSA samples, and the solid ones to two PANI-HCl samples that exhibit $\alpha=1$ below RT. The lines are guides to the eye. The uppermost symbols correspond to $T^*$, whereas the bottom symbols show $T_{\mathrm{LOW}}$ which is the lowest temperature at which $\sigma$ was measured. The intermediate symbol for the two highest $\eta_1$ values mark crossover between $\alpha=2/5$ and $\alpha=1/4$. A sketch of the theoretically predicted diagram \cite{fts} is shown in the inset.}
\end{center}
\end{figure}

From the above analysis, it is rather clear that several predictions of the FTS model agree with our experimental results. The remaining issue is whether we can construct the FTS diagram - the proposed shape of which is sketched in the inset to Fig.~\ref{fig3} - from our data. This diagram maps areas of different $\alpha$  in a $T$ versus disorder level plane. VRH exponents 1/4, 2/5, and 1/2 can all appear only for $d=3$ \cite{fts}, and this dimensionality should hence apply to our samples. Since data on crossovers  between different $\alpha$ are available from our $\sigma(T)$, constructing the FTS diagram reduces to parameterizing the disorder level. This is, however, not straightforward. Disorder in conducting polymers arises from a combination of molecular-scale disorder (see e.g. \cite{hPS} and references therein) and structural inhomogeneities at mesoscopic scale lengths \cite{KaiserAM}. Thus, information on averaged structural ordering (e.g., from X-ray diffraction) can be instructive but not conclusive in inferring what level of disorder a hopping electron experiences, particularly in the case of our PANI-DBSA samples which are fully amorphous. Having this in mind, we approach the problem by using a macroscopic property of $\sigma$ as a parameter accounting for disorder.

\bigskip

We believe that $\eta_{\alpha}$   for  $\alpha=1$, i.e., experimental $\eta_1$, can be used as an appropriate measure of the disorder level in our samples. This plausible assumption, as we show later, leads to our results reproducing the FTS diagram, and the reasoning behind choosing $\eta_1$  is as follows.  Generally, the exponential part in Eq.~(\ref{equation1}) pertains mainly to thermal aspects of  $\sigma$  while  $\eta_{\alpha}$  is related to non-thermal effects - in which disorder plays a crucial role. Since  $\sigma \propto \eta_{\alpha}$, as disorder level decreases, $\eta_{\alpha}$ should increase. This holds although  $\eta_{\alpha}$ also contains a contribution of the charge injected by the doping. Protonation not only increases the charge-carrier density (which also affects the screening of the Coulomb interaction and thusly influences $T_{\alpha}$ \cite{mario}) but also reduces disorder in the hopping landscape \cite{Pouget1991}. This occurs because the Coulomb interaction energy between dopant anions - which are bound to the protonated sites together with the potential wells - is lowered in a more ordered configuration. Another possible effect of adding a dopant is enhancement of the interchain hopping \cite{cca}, which is also a process that homogenizes the potential. Of all $\eta_\alpha$, we focus on $\eta_1$ because it accounts for the described processes in the same way for all the samples which exhibit NNH, regardless of the low-$T$ value(s) of $\alpha$.   

\bigskip

In Fig.~\ref{fig3}, we plot areas of different values of $\alpha$  in a $(\eta_1,T)$ plane, as indicated.  Each $\eta_1$ value is related to one sample, and the symbols have been extracted from the corresponding $\sigma(T)$.  The vertical distance between two symbols for a given $\eta_1$ is the $T$ range of the appearance of the indicated VRH $\alpha$. The open symbols correspond to the PANI-DBSA samples, and the solid symbols to the only two of our PANI-HCl samples (both with $Q\lesssim2$)  that exhibit well defined $T$ ranges of $\alpha=1$ below RT (their undoped PANI being the same as for the PANI-DBSA samples). The uppermost symbols correspond to $T^*$ (the error bars indicate the crossover widths), whereas the bottom symbols represent the lowest temperature $T_{\mathrm{LOW}}$ at which $\sigma$ was measured (the low-$T$ limit of the cryogenic system, or $T$ at which $\sigma$ is immeasurably small). For the two samples with the highest $\eta_1$, the intermediate symbol represents crossover between $\alpha=2/5$ and $\alpha=1/4$. The dashed lines are guides to the eye, depicting characteristic temperatures extracted from  $\sigma(T)$, whereas the dotted lines mimic other lines from the theoretical diagram.

\bigskip

The similarity of the experimental and theoretically predicted diagrams is obvious. Furthermore, the two PANI-HCl samples show the same behavior as the PANI-DBSA samples. We believe that this is not a coincidence, because the starting undoped PANI has been the same for both classes of samples and every processing (set by the choice of the dopant and the doping method) modified the same starting material. On the other hand, a particular processing leaves a distinct signature in $\eta_1$. For instance, both PANI-HCl samples are almost fully protonated but they were processed differently and hence do not display the same properties \cite{jncs}. The sample with $\eta_1 \sim 2000$ S/m exhibits $\alpha=1/4$ while this regime is absent for the sample with $\eta_1 \sim 1100$ S/m. Moreover, the PANI-HCl samples are partly crystalline \cite{mario} while the PANI-DBSA samples are 
amorphous, which additionally supports the conclusion that degree of crystallinity is not a proper measure of the (dis)order in a potential that hopping electrons experience. It seems, however, that the doping-induced ordering of the hopping potential must be strong enough for different classes of doped PANI to show the same HET behavior quantitatively, since our less ordered PANI-HCl samples from the same starting material do not exhibit NNH below RT \cite{mario}. That is, the $\eta_1 < 1000$ S/m parts of the FTS diagrams for PANI-HCl and PANI-DBSA differ quantitatively although the corresponding VRH properties are 
similar \cite{mario}. In contrast to PANI-HCl, $T^*$ for PANI-DBSA does not increase above RT until $\eta_1$ decreases below $\sim 75$ S/m. This increase is depicted by the dotted line for $\eta_1 \to 0$, which can be drawn thusly because there is no NNH below RT for the least conducting of the PANI-DBSA samples (see Fig.~\ref{fig1}).

\bigskip

In developing their model, FTS primarily had in mind systems such as stripe phases, quantum wire arrays, carbon-nanotube films, compounds like Bechgaard salts and Fabre salts, etc. They also mentioned \cite{Teber2006,fts} conducting polymers as candidates since VRH has been reported extensively for these materials. Yet, FTS were seemingly not fully confident in the applicability of their model in this case, possibly because of the intricate underlying morphology. It turns out, however, that the model is robust enough to account for these complex HET conductors as well.

\bigskip

In conclusion, the $\sigma(T)$ curves (between 10 K and RT) of our doped PANI pellets reveal both VRH and NNH. This is observed for doping with DBSA over a wide range as well as for nearly full protonation by HCl. The two HET mechanisms extend over $T$ ranges that are sufficiently large to permit studying of the full complexity of the FTS diagram. We determine crossover temperatures between regimes of different HET exponents. When we plot these temperatures against a measure of the disorder level, that is, prefactor of the NNH Arrhenius plots, we obtain a remarkable agreement with the predicted shape of the FTS diagram. This implies that the FTS model is both correct and applicable to conducting polymers.

\vspace{20pt}

%%%%%%%
\ack%%%
%%%%%%%

This work has been funded by the Croatian Ministry of Science, Education and Sports under grant No. 119-1191458-1008.

\vspace{25pt}

%\begin{acknowledgments}
%\end{acknowledgments}

%%%%%%%%%%%%%%%%%%%%%%%%%%%%%
\section*{References}%%%%%%%%
%%%%%%%%%%%%%%%%%%%%%%%%%%%%%
%
\bibliography{iop}% Produces the bibliography via BibTeX.

\end{document}